\documentclass[authoryear,10pt]{article}
\usepackage[a4paper]{geometry}
\usepackage{graphicx}
\usepackage[textwidth=8em,textsize=small]{todonotes}
\usepackage{amsmath}
\usepackage{natbib}

\usepackage{placeins}

\usepackage[hyphens]{url}
\usepackage{hyperref}
\usepackage[hyphenbreaks]{breakurl}

\begin{document}

\title{Preventing the Forecaster's Evaluation Dilemma}
\author{Malte C. Tichy \\ 
Blue Yonder GmbH, Oberbaumbrücke 1, 20457 Hamburg, Germany}

\maketitle

\begin{abstract}
Assume that a grocery item is sold 1'234 times on a given day. What should an ideal forecast have predicted for such a well-selling item, on average? More generally, when considering a given outcome value, should the empirical average of forecasted expectation values for that outcome ideally match it? Many people will intuitively answer the first question with ``1'234, of course'', and affirm the second. Perhaps surprisingly, such grouping of data by outcome induces a bias in the evaluation. An evaluation procedure that is aimed at verifying the absence of bias across velocities, when based on such segregation by outcome, therefore fools forecast evaluators and incentivizes forecasters to produce overly exaggerated (extreme) forecasts. Such anticipatory adjustments jeopardize forecast calibration and clearly worsen the forecast quality -- this problem was named the \emph{Forecaster's Dilemma} by Lerch {\it et al.} in 2017 (Statististical Science {\bf 32}, 106). As a solution to check for bias across velocities, forecast evaluators should group pairs of forecasts and outcomes by the predicted values, and evaluate empirical mean outcomes per prediction bucket. Within a simple mathematical treatment for the number of items sold in a supermarket, the reader is walked through the dilemma and its circumvention.
\end{abstract}

\section{Introduction: Evaluation-induced biases}
Forecasts for categorical, continuous, or countable events can be evaluated as soon as the event under interest has occurred, and the actual category, fractional number, or integer value is revealed. Such forecast evaluations are not always systematically performed in an unbiased fashion, but they are often motivated by certain exterior circumstances: As long as the process that is steered by some forecast runs smoothly, e.g., the replenishment of fresh strawberries in a supermarket leads to happy buyers, there is no reason to question forecast quality. When an extreme event occurs and the process is jeopardized (strawberries are out of stock already at mid-day, with nefarious consequences on customer satisfaction), the question is quickly asked whether that extreme event had been predicted well.  \cite{LerchForecastersDilemma} point out that such an  event-driven way to select data induces a bias in the evaluation: Extreme events then appear to have been predicted to be less extreme than observed, even if the forecast is calibrated in the sense of \citep{probforecastingcalib}. On the one hand, this misguides forecast evaluators, who might wrongly accuse calibrated forecasts to be of little predictive value. On the other hand, it incentivizes forecasters to dishonestly exaggerate the probability of extreme events and fabricate a conspiratory forecast. This conflict of interests is clearly problematic, motivating \cite{LerchForecastersDilemma} to designate it as \emph{Forecaster's Dilemma}. 

When a forecaster knows to be only evaluated when extreme events (financial crises, surprising outcomes of elections, catastrophes etc.) occur, they have the incentive to exaggerate the probability of such events, even if such anticipatory adjustments jeopardize the calibration of their forecast. As long as the predicted extreme events do not occur, the forecast is not being evaluated and no harm is done. If the extreme event materializes, the forecast evaluation will be benevolent, with great personal benefit. For example, predicting a financial crisis for the next year remains unnoticed as long as the economy is doing well, and a doomsayer does not need to expect any negative consequences induced by the pessimism they express. The same forecast then guarantees attention, live TV interviews, well-payed contracts, and book sales if the crisis actually breaks out. Clearly, a meaningful forecast is an honest and calibrated one: A financial crisis should occur in about 10\% of the years for which one predicted a financial crisis to occur with 10\% chance  if that forecast is calibrated. 

These notes expand the argument of \cite{hindsight_bias_blog} and constitute a technical addendum to \cite{forecastersdilemma_foresight}; to make the text self-contained, I closely repeat the lines of thought put forward in these publications. I show that even the seemingly straightforward segregation of prediction-outcome-pairs by outcome induces a bias in the evaluation. The recommendation to the reader in their assumed role as a forecast evaluator is to group a collection of forecasts by prediction and to average the outcomes per prediction value (or by buckets of nearby predictions, since predictions for expected values are typically continuous, even for discrete outcomes). This procedure then reveals whether the forecast is at least calibrated in the sense that predicted expectation values can be trusted across all ranges (from slow to fast). Communicating this procedure to the forecast creator incentivizes them to provide calibrated forecasts and to not exaggerate the probability of extremes. 

\section{Prediction setup}
We consider the demand predictions for grocery items for a single day at a retailer that offers a large number $n \sim 10'000$ of different items. All items are sold by piece (i.e., not by kg or liter). The number of sold pieces of every item is governed by a selling rate $r$. The forecast produces $n$ rates $r_1$ to $r_n$. Formally, the forecast draws out of a continuous probability distribution $P_\text{rate}(r)$, which characterizes the entire assortment. The rate $r_j$ predicts the expectation value of the number of sales of item $j$. The sales process is assumed to be Poissonian  \citep{Tichy_tech_2023}, that is, the actual sales $s_j$ are believed to fulfill 
\begin{equation}
s_j \sim P_{\text{process}}( s=s_j | r=r_j) = P_{\text{Poisson}}( s=s_j | r=r_j) = \frac{r_j^{s_j}}{s_j!}e^{- r_j} . \label{poissonprediction}
\end{equation}
We use the current situation as an illustratory example. The overall argument is, however, independent of the chosen distributions and processes. 

Our focus lies on the evaluation procedure. For this purpose, we will assume that the forecast $r_1 \dots r_n$ is calibrated, that is, that the sales $s_j$ are truly distributed according to the prediction Eq.~(\ref{poissonprediction}). Any ostensible bias that the evaluation reveals is known to not be a property of the forecast, but induced by the evaluation procedure.

\section{Evaluation setup}
Before possessing a specific forecast $r_j$ for the item $j$, that is, before being given any information on the specific selling rate of that item, we already know that the target is distributed following the prior,
\begin{equation} 
s_j \sim  P_{\text{target}}(s)= \int \text{d}r P_\text{rate}(r) P_{\text{process}}(s|r)  .
\end{equation}

That is, the probabilistic forecast $P_{\text{target}}(s)$ (issued every time a forecast is being asked for) is very unsharp, yet it is calibrated: We can expect, in the long term, the empirical distribution of $s$ to match the predicted one. It does not, however, individualize at all for different items $j$. On the other hand, $s_j \sim P_{\text{process}} (s_j|r_j) = P_{\text{Poisson}} (s_j|r_j)$ (with a certain value of $r_j$ provided for the item of interest) is the sharpest possible forecast if we accept that the Poissonian uncertainty of the probabilistic process is insurmountable \citep{Tichy_tech_2023}. 

In general, there can be forecasts of intermediate sharpness, for which the uncertainty in the Poisson rate for a given item $j$ is neither described by the distribution $P_{\text{rate}}(r)$, nor by a delta-function, but by some distribution of intermediate character. For example, if the forecaster's knowledge of the rate is expressed by the gamma distribution with mean $\rho$ (and some implicit shape parameter), the resulting 
\begin{equation}
P(s | \rho )= \int \text{d}r P_\gamma (r | \rho ) P_{\text{process}} (s|r)
\end{equation}
becomes negative-binomial in shape. 

The overall mean observation and the overall mean forecast naturally match, reflecting the ask that a forecast be globally unbiased: 
\begin{equation}
E(s) = \sum_{s=0}^{\infty}  s P_{\text{target}} (s)   = \sum_{s=0}^\infty s \int \text{d}r  P_{\text{rate}} (r)  \\ P_{\text{process}} (s|r) = \int \text{d}r  P_{\text{rate}} (r) r = E(r) \label{overallbias}
\end{equation}

That is, a forecast evaluator can compare the empirical values for the overall mean predictions and outcomes, 
\begin{eqnarray}
\bar r &=& \frac{1} n \sum_{j=1}^{n} r_j , \label{overallmeanprediction} \\
\bar s &=& \frac{1} n \sum_{j=1}^{n} s_j ,  \label{overallmeanobservation} 
\end{eqnarray}
and test whether any discrepancy between them is statistically significant.

\subsection{Velocity-dependent bias}
Verifying the property of Eq.~(\ref{overallbias}) assures the evaluator that a forecast is globally unbiased: The overall mean prediction in Eq.~(\ref{overallmeanprediction}) and the mean observation in Eq.~(\ref{overallmeanobservation}) match (within the expected statistical margin). This property is, however, certainly not sufficient to verify that the forecast is meaningful: Even a forecast in which the true rates are randomly reassigned to other items (map any $r_j$ to $r_{p(j)}$ with $p$ being some permutation of $\{1, \dots, n\}$) would fulfill that property, and so does the most unsharp forecast that produces $r_j=\bar r$ for every item $j$.  As a first step towards verifying the individualization of the forecast among the items $j=1,\dots, n$, one can tackle the question of whether the forecast exhibits any velocity-dependent bias, i.e.~whether the ``fast movers'' (which sell many times per day) are just as unbiased as the ``slow-movers'' (which only sell few times or even less per day). It is, for example, possible that the fast-movers are systematically over-predicted and the slow-movers systematically under-predicted (or vice-versa), which could result in a vanishing global bias, but a strong velocity-dependent bias. In other words, given $n$ pairs of predictions and outcomes $(r_1, s_1)$, $(r_2, s_2)$, $\dots,$ $(r_n, s_n)$, instead of globally comparing $\bar r$ to $\bar s$, we would compare $\bar r$ to $\bar s$ \emph{within certain groups of forecasts}.

The question arises how to define such velocity groups: Shall one ask for the mean prediction $r$ per observation value $s$, or for the mean observation $s$ per prediction $r$ (or interval of predictions $r - \frac{\epsilon}{2} < R < r + \frac{\epsilon}{2} $)? The first stance is backward-looking, since it asks ``What was the forecast, given a certain outcome?'', while the second is forward-looking, asking ``What will the outcome be, given a certain forecast?''.

Given the data, one can compute the mean observation $ \bar s^{({r})}$ per $\epsilon$-interval around the prediction $r$, 
\begin{equation}
\bar s^{({r})} = \frac{\sum_{k=1}^n s_k  \cdot \delta_{r - \frac{\epsilon}{2 } < r_k < r + \frac{\epsilon}{2 }}  }{   \sum_{k=1}^n \delta_{r - \frac{\epsilon}{2 } < r_k < r + \frac{\epsilon}{2 }}   }  ,  \label{empiricalsum_s}
\end{equation}
where the sums include all items $k$ such that $r_k$ is in the interval around $r$. 
Likewise, one can compute the mean prediction $ \bar r^{({s})}$ per observation $s$,
\begin{equation}
\bar r^{({s})} = \frac{\sum_{k=1}^n r_k \cdot \delta_{s_k=s}  }{   \sum_{k=1}^n \delta_{s_k=s}  }  . \label{empiricalsum_r}
\end{equation}
In practice, the often unquestioned expectations are that a perfect forecast should fulfill 
\begin{eqnarray}
\bar r^{({s})} & \approx & s   \label{empirical_r_given_s}, \text{ for all observed values }s ,  \label{expectation_backward} \\
\bar s^{({r})} &\approx & r  \label{empirical_s_given_r}, \text{ for all predicted values }r   ,
\label{expectation_forward}
\end{eqnarray}
at least when the resulting groups over which the sums in Eqs.~(\ref{empiricalsum_s}, \ref{empiricalsum_r}) are generated are sufficiently large. 

To many practitioners, the expectation expressed by Eq.~(\ref{expectation_backward}) seems more ``natural'' and intuitive than the other one of Eq.~(\ref{expectation_forward}), because one thereby groups the data by the real, actual outcomes, instead of grouping by more ``fugitive'' forecasts. Let us now put these often implicit expectations to the test.

\subsection{Forward-looking evaluation, grouped by prediction}
In terms of probabilities, Eq.~(\ref{empirical_s_given_r}) is the empirical realization of the conditional expectation value of the outcome $s$, given predictions being in the interval around $r$. We have, by construction,
\begin{equation}
E(s|r)= \sum_{s=0}^\infty  s P_{\text{process}} (s|r)= r ,  \label{forwardlexpvalue}
\end{equation}
which summarizes the forecast’s calibration promise: Averaging the outcomes $s$ for all the forecasts that produced the expectation value $r$, we should find $r$ as a result. Empirically, we can indeed expect that Eq.~(\ref{empirical_s_given_r}) holds, in the sense that this equality is not broken in a statistically significant way.

Eq.~(\ref{forwardlexpvalue}) confirms that it is useful to ask the forward-looking question and to compare the empirical realization $\bar s^{({r})}$ of $E(s|r)$ to the expected mean $r$. This can be done in buckets of predictions of different sizes, where a balance has to be established between differentiation (more and smaller intervals) and statistical significance (fewer and larger intervals). 

\subsection{Backward-looking evaluation, grouped by outcome} \label{backwardlooking}
The backward-looking question concerns the conditional expectation value of the prediction $r$, given the outcome $s$,
\begin{equation}
E(r|s) = \int \text{d}r ~r \cdot P_{\text{hindsight}} (r | s)  ,
\end{equation}
where the reversed conditional probability $P_{\text{hindsight}} (r | s)$ describes the hindsight probability to \emph{have had predicted} $r$, \emph{after having observed} $s$. Via Bayes’ rule, we can relate it to the process, the rate distribution, and the target distribution via
\begin{equation}
P_{\text{hindsight}} (r|s)= \frac{ P_{\text{process}} (s|r) P_{\text{rate}} (r) }{P_{\text{target}} (s)} , \label{hindsightbayes}
\end{equation}
and we find
\begin{eqnarray}
E(r|s)&=&  \frac 1 {P_{\text{target}} (s)}  \int \text{d}r ~r \cdot  P_{\text{process}} (s|r) P_{\text{rate}} (r) \label{hindsightexpvalue} \\
&=& \frac{\int \text{d}r ~r \cdot  P_{\text{process}} (s|r) P_{\text{rate}} (r)}{\int \text{d}r  P_{\text{process}} (s|r) P_{\text{rate}} (r) }  
\end{eqnarray}
This expression looks intricate, and there is no general reason why it should always evaluate to the value $s$. That is, the expectation inscribed in Eq.~(\ref{empirical_r_given_s}) has no corroboration. Forecasters should not aim at producing a set of forecasts that fulfills Eq.~(\ref{empirical_r_given_s}), but dismiss that ask, and suggest testing for calibration instead. Forecast evaluators should be careful when interpreting the empirical values $\bar r^{(s)}$, since their ideal expectation value has the intricate form given by Eq.~(\ref{hindsightexpvalue}). 

What the forecast \emph{should have had predicted} on average for a certain outcome $s$ depends not only on the process $P_{\text{process}} (s|r)$, but also on the entire a priori distribution of rates $P_{\text{rate}} (r)$. All rates $r$ can, in principle, feed an outcome $s$, which is why every value of $r$ contributes, weighted by its a priori probability $P_{\text{rate}} (r)$ and by its probability to have fed $s$ $P_{\text{process}} (s|r)$. What is reasonable to have predicted for a certain outcome $s$ heavily depends on the context, set by the entire distribution $P_{\text{rate}} (r)$: Ten sold luxury cars for a small car dealer have likely been the best-selling day in a decade, and severely under-predicted, whereas ten bottles of orange juice in a supermarket have probably been a disappointing sale, and strongly over-predicted. Depending on the shape of $P_{\text{rate}} (r)$, the contributions of large or small values of $r$ will push the expectation value $E(r|s)$ to large or small values. Assuming a finite support of $P_{\text{rate}} (r)$, large values of $s$ will have more contributions from $r < s$, and small values of $s$ will be rather fed by $r > s$. This means that choosing a large outcome $s$ typically means choosing an under-forecasted one, and choosing a small outcome $s$ means choosing an over-forecasted one.

The case $s=0$ is particularly insightful: It is only fed by rates $r>0$. There is no such thing as a negative sales prediction. All items that were not sold on a given day were unavoidably over-forecasted -- if the forecast evaluator accepts that a forecast is probabilistic, they also need to accept the ostensible ``bias'' reflected by $E(r|s)\neq s$.

\section{Conclusions}
For the successful usage of forecasts in practice, forecast creators and forecast users should  align on expectations, assumptions, and limitations. For example, a forecast user should be transparent on how they evaluate the forecast \citep{Kolassa2020}, i.e., they should disclose whether they use Mean Absolute Error or some other function of the outcome and a forecasted point estimate, or a procedure tailored to the probabilistic nature of their forecast. Depending on the business usage of the forecast, one metric is more suitable than the other, since the metric should reflect the true business cost of the deviation between the forecasted point estimate and the actual. Revealing the metric allows the forecaster to extract the optimal point estimate for that metric out of their predicted probability distribution. This is a perfectly honest and meaningful procedure: The ``best'' forecasted point value depends on the chosen evaluation metric, paraphrasing the title of \citep{Kolassa2020}. What is ``best'' is a matter of perspective. The situation treated in this short contribution is totally different: The very expectation that the mean of the predictions for a certain outcome value should match that outcome, inscribed in Eq.~(\ref{expectation_backward}), is flawed and needs to be abandoned. For forecaster creators, there is no way around educating forecast evaluators on the perhaps unexpected properties of Eqs.~(\ref{hindsightbayes},\ref{hindsightexpvalue}), and guide the evaluation towards a meaningful one, based on Eq.~(\ref{forwardlexpvalue}).

Many different a priori distributions of the rate $P_{\text{rate}}(r)$ and all kinds of processes $P_{\text{process}} (s|r)$ are thinkable, such that no strong general statements can be made about the values that $E(r|s)$ takes. This reflects the unsurprising fact that the question ``Which forecast made sense for this outcome?'' is extremely context-dependent, and hinges on the prior expectation: 10 degrees celsius at noon in January in Oslo (Norway) will have been underforecasted, the same temperature in July in Dallas (Texas) has probably been overforecasted. It is natural to jump to select the best- and worst-selling items and check their forecast; we must let go of the expectation, however, that a good forecast should have predicted these in an ``unbiased'' fashion: By our very way of selecting the data, we have induced a bias, which overshadows any possible velocity-dependent forecast bias.

\end{document}